\documentclass[12pt]{article}
\usepackage{epsf}
\title{ {\bf
The scalar unparticle effect on the charged lepton electric dipole
moment}}
\author{\vspace{1cm}\\
        {\bf E. O. Iltan}
        \thanks{E-mail address:
        eiltan@newton.physics.metu.edu.tr}
 \\
        Physics Department, Middle East Technical University \\
        Ankara, Turkey\\}

\date{}

\begin{document}
\setlength{\baselineskip}{24pt}
\maketitle
\setlength{\baselineskip}{7mm}
\begin{abstract}
We study the charged lepton electric dipole moment which is
induced by the scalar unparticle mediation and we predict the
appropriate range for the free parameters appearing in the
effective lagrangian which drives the unparticle-standard model
lepton interactions. We observe that the charged lepton electric
dipole moment is strongly sensitive to the scaling dimension $d_u$
of the unparticle and the new couplings in the effective
interaction. Furthermore, we see that the current experimental
limits of charged lepton electric dipole moments can ensure an
appropriate range for these free parameters.
\end{abstract}
\thispagestyle{empty}
\newpage
\setcounter{page}{1}
\section{Introduction}
Scale invariance has been studied in various disciplines of
physics and the scale invariant field theories has been
investigated extensively. However, it is not clear that how such
possible scale invariant sector comes out in the experiments and
how it is determined. Recently, Georgi \cite{Georgi1} proposed a
hypothetical non-trivial scale invariant sector, having a
non-trivial infrared fixed point. It is a new one beyond the
standard model (SM) and has the scaling dimension $d_u$. Georgi
suggest that at low energy, around $\Lambda_U\sim 1\,TeV$, this
sector appears as so called unparticle stuff with non-integer
scaling dimension $d_u$ and it looks like a number of $d_u$
massless particles which are invisible. The properties of the
suggested matter, the unparticle, are different from the
properties of the particle \cite{Georgi1,Georgi2}. The unparticles
couple to the SM fields and the interactions are described by the
effective lagrangian in the low energy level. This approach makes
it possible to test the effects of the possible scale invariant
sector experimentally with the help of the unparticle idea. The
missing energies at various processes which can be measured at LHC
or $e^+e^-$ colliders, the various decays and the dipole moments
of fundamental particles (that can occur with the unparticle
mediation theoretically) are among the possible candidates to
search the unparticles and their properties. There is an extensive
phenomenological work done in the literature to search the
possible effects of the unparticles
\cite{Georgi2}-\cite{Balantekin}: the effects on the missing
energy of many processes, the anomalous magnetic moments,
$D^0-\bar{D}^0$ and $B^0-\bar{B}^0$ mixing, lepton flavor
violating interactions, direct CP violation in particle physics;
the phenomenological implications in cosmology and in
astrophysics.

In this work we study the contribution of the scalar unparticle to
the electric dipole moments (EDMs) of the charged leptons. The
fermion EDMs are worthwhile to examine since their existence is
the sign for the CP violation which is among the most interesting
physical phenomena. There are number of experimental and
theoretical studies on fermion EDMs in the literature. The current
experimental results on the lepton EDMs are, $d_e =(1.8\pm 1.2\pm
1.0)\times 10^{-27} e\, cm$ \cite{Commins}, $d_{\mu} =(3.7\pm
3.4)\times 10^{-19} e\, cm$ \cite{Bailey} and $d_{\tau}
=(3.1)\times 10^{-16} e\, cm$ \cite{Groom}, respectively. From the
theoretical point of view, for non-zero fermion EDMs, one needs a
CP violating phase, which is carried by the complex Cabibo
Kobayashi Maskawa (CKM) (the lepton mixing matrix) in the quark
(lepton) sector, in the SM. Since the theoretical values of the
fermion EDMs in the SM are negligibly small (see for example
\cite{Khiplovich1}-\cite{Czarnecki2}), one investigates new models
beyond, in order to obtain the numerical values not far from the
experimental limits. The lepton EDMs have been studied
theoretically in various models \cite {Bhaskar}-\cite
{IltanRSEDM}; in the framework of the seesaw model \cite
{Bhaskar}, in the two Higgs doublet model (2HDM) \cite
{Iltmuegam}, in the framework of the SM with the inclusion of
non-commutative geometry \cite{IltanNonCom}, with the inclusion of
non-universal extra dimensions \cite{IltanExtrEDM}, in the split
fermion scenario \cite{IltanSplitEDM, IltanSplitHiggsLocalEDM}, in
the Randall Sundrum background \cite{IltanRSEDM}.

In the present work, we consider that the scalar unparticle
mediation is responsible for the CP violating interaction which
leads to the charged lepton EDM and we predict the appropriate
range for the free parameters appearing in the effective
lagrangian which drive the unparticle-SM lepton interactions. We
observe that the charged lepton EDMs are strongly sensitive to the
scaling dimension $d_u$ and the new unparticle-SM lepton
couplings. Furthermore, the current experimental limits of charged
lepton EDMs can ensure an appropriate range for these free
parameters.

The paper is organized as follows: In Section 2, we present the
contribution of the scalar unparticle to the charged lepton EDMs.
Section 3 is devoted to discussion and our conclusions.
\section{Electric dipole moments of charged leptons induced by
the scalar unparticle}
The fermion EDM is among the most powerful tools to search the CP
violation since it is driven by the CP violating
fermion-fermion-photon effective interaction. The CP violation is
carried by the complex phase of the CKM matrix (a possible lepton
mixing matrix) in the quark (lepton) sector, in the SM. However,
their theoretical values are the extremely small in the SM since
they exist at beyond two loop orders and the predictions in this
model are far from the current experimental limits (see for
example \cite{Khiplovich1}-\cite{Czarnecki2}). In the present
work, we consider that the unparticle stuff is also responsible
for the CP violating interaction which stimulates the lepton EDMs.
The unparticle idea has been created by Georgi \cite{Georgi1,
Georgi2} and this idea is based on the very high energy level
interaction of two sectors, the SM and the ultraviolet sector with
non-trivial infrared fixed point. Around $\Lambda_U\sim 1\,TeV$
this sector manifest itself as new degrees of freedom, called
unparticles, being massless and having non integral scaling
dimension $d_u$. Finally, at low energies the related interactions
might be driven by the effective field theory and the
corresponding effective Lagrangian might be
\begin{equation}
{\cal{L}}_{eff}\sim \frac{\eta}{\Lambda_U^{d_{tot}}}\,O_{SM}\,
O_{U} \,, \label{efflag}
\end{equation}
where $d_{tot}=d_u+d_{SM}-n$ with the space-time dimension $n$.
Here the parameter $\eta$ is related to the energy scale of
ultraviolet sector, the low energy one and the matching
coefficient (see \cite{Georgi1,Georgi2,Zwicky} for details). Since
the operators with lowest possible dimension have the most
powerful effect in the low energy effective theory, one chooses
the appropriate operators $O_U$ which have all possible Lorentz
structures (see for example \cite{SChen}).

Our aim is to study the charged lepton EDMs and we need the
effective interaction among the unparticle and the SM leptons in
order to construct the dipole type transition. The lowest order
effective interaction we are interested in is:
\begin{eqnarray}
{\cal{L}}= \frac{\eta^{S}_{ij}}{\Lambda^{du-1}} \bar{l}_{i L}
\,l_{j R}\,O_{U} + h.c. \,\,\, , \label{lagrangian1}
\end{eqnarray}
where $L$ and $R$ denote chiral projections $L(R)=1/2(1\mp
\gamma_5)$, $l$ is the lepton field,  $O_{U}$ is the scalar
unparticle operator. The couplings $\eta^{S}_{ij}$  are free
parameters, complex in general and the scalar (pseudoscalar)
$\lambda_{ij}^{S}$ ($\lambda_{ij}^{P}$) coupling reads
\begin{eqnarray}
\lambda_{ij}^{S(P)}=\frac{1}{2}\Big(\eta_{ij}^{S}\pm
(\eta_{ij}^{S})^*\Big) \, . \label{lambdaij}
\end{eqnarray}
This is the case that a possible new source for the CP violating
interaction emerges (see Discussion section) and the EDM of the
charged leptons due to the unparticle stuff, here the scalar
unparticles, appear.

\begin{figure}[htb]
\vskip -5.5truein \hspace{6cm} \epsfxsize=4.8in
\leavevmode\epsffile{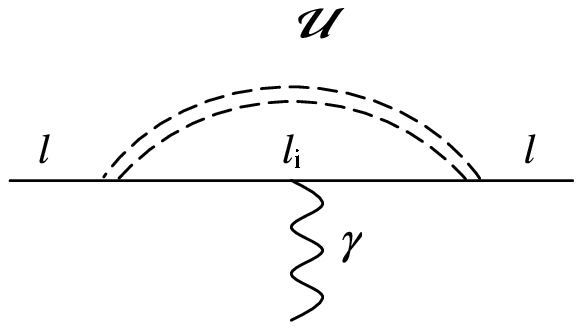} \vskip 8.5truein \caption[]{One loop
diagram contributes to the EDM of charged leptons due to scalar
unparticle. Wavy (solid) line represents the electromagnetic field
(lepton field) and double dashed line the scalar unparticle
field.} \label{fig1}
\end{figure}

The effective EDM interaction for a charged lepton $l$ is
\begin{eqnarray}
{\cal L}_{EDM}=i d_l \,\bar{l}\,\gamma_5 \,\sigma^{\mu\nu}\,l\,
F_{\mu\nu} \,\, , \label{EDM1}
\end{eqnarray}
where $F_{\mu\nu}$ is the electromagnetic field tensor, '$d_{l}$'
is the EDM of the charged lepton which is a real number by
hermiticity. The possible CKM type lepton mixing matrix causes the
non-zero EDMs for the charged leptons, however, here, we consider
only the effective interaction of the SM leptons and the
unparticle operator to create the CP violation, in order to make
predictions of the acceptable range for the free parameters of the
unparticle stuff, by using the experimental current limits. From
the theoretical point of view, the lepton EDM arises at least in
the one loop level with the mediation of the scalar unparticle, in
the present case (see Fig.\ref{fig1}). The scalar unparticle
propagator is obtained with the help of the scale invariance and
using the two point function of the unparticle, it is obtained as
\cite{Georgi2, Cheung1}
\begin{eqnarray}
\int\,d^4x\,
e^{ipx}\,<0|T\Big(O_U(x)\,O_U(0)\Big)0>=i\frac{A_{d_u}}{2\,\pi}\,
\int_0^{\infty}\,ds\,\frac{s^{d_u-2}}{p^2-s+i\epsilon}=i\,\frac{A_{d_u}}
{2\,sin\,(d_u\pi)}\,(-p^2-i\epsilon)^{d_u-2} \, ,
\label{propagator}
\end{eqnarray}
with
\begin{eqnarray}
A_{d_u}=\frac{16\,\pi^{5/2}}{(2\,\pi)^{2\,d_u}}\,\frac{\Gamma(d_u+\frac{1}{2})}
{\Gamma(d_u-1)\,\Gamma(2\,d_u)} \, . \label{Adu}
\end{eqnarray}
For $p^2>0$, the function $\frac{1}{(-p^2-i\epsilon)^{2-d_u}}$ in
eq. (\ref{propagator}) becomes
\begin{eqnarray}
\frac{1}{(-p^2-i\epsilon)^{2-d_u}}\rightarrow
\frac{e^{-i\,d_u\,\pi}}{(p^2)^{2-d_u}}
\, , \label{strongphase}
\end{eqnarray}
with a non-trivial phase as a result of non-integral scaling
dimension.

Now, we would like to present the EDMs of charged leptons $l$,
driven by the scalar unparticle mediation:
\begin{eqnarray}
d_l=
-i\frac{e}{32\,\pi^2}\frac{A_{d_u}}{\Lambda^{2\,d_u-2}\,sin\,(d_u\pi)}
\,\sum_{i=1}^3\,\Big\{\lambda_{il}^S\,((\lambda_{il}^P)^*-\lambda_{il}^P)\,
Q_i\,m_i
\int^{1}_{0}\,dx\,\int^{1-x}_{0}\,dy\,f(x,y,m_i,m_l)\Big\} ,
\label{EDMscalar}
\end{eqnarray}
where subscript $i$ denotes the internal charged leptons,
$i=e,\mu, \tau$, $Q_i$ is the charge of $i^{th}$ lepton and
$f_1(x,y,m_i,m_l)$ is the function given by
\begin{eqnarray}
f(x,y,m_i,m_l)=\frac{(1-x-y)^{1-d_u}\,(x+y)}{\Big( (x+y)\,
(m_l^2\,(1-x-y)-m^2_i)\Big)^{2-d_u}}\, . \label{f1}
\end{eqnarray}
Notice that both scalar and pseudoscalar couplings of unparticles
to leptons should be considered to obtain  non-zero EDM.
%
\section{Discussion}
This section is devoted to the analysis of the possible effects of
the scalar unparticle on the charged lepton EDMs and the
prediction of the scale dimension of the unparticle by considering
the current experimental limits of the EDMs. The non-zero EDM of a
fermion is the evidence for the existence of the CP violation
since it is driven by the CP violating fermion-fermion-photon
effective interaction. The source of the CP violating phase in the
SM is the complexity of the CKM matrix in the quark sector and the
possible lepton mixing matrix in the lepton sector with non-zero
neutrino masses. In the present work, we assume that there exist a
new CP phase that arises from the complex couplings, appearing in
the tree level fermion-fermion-unparticle interaction, in the
effective theory and we investigate its contribution to the
charged lepton EDMs. Here the couplings, the CP phase and the
scale dimension of the unparticle(s) are the free parameters which
should be restricted.

\begin{itemize}
\item The scale dimension $d_u$ is restricted in the range $1< d_u
<2$. Here,  $d_u>1$ is due to the non-integrable singularities in
the decay rate \cite{Georgi2} and $d_u<2$ is due to the
convergence of the integrals \cite{Liao1}.
\item Here, we take the coupling $\eta^S_{ij}$ in
eq.(\ref{lagrangian1}) as $\eta^S_{ij}=e^{i\theta}\,|\eta^S_{ij}|$
where $\theta$ is the CP violating parameter. In terms of
$\eta^S_{ij}$ the scalar and pseudoscalar couplings read
$\lambda_{ij}^{S}=\frac{1}{2}(\eta^S_{ij}+(\eta^S)^*_{ij})$ and
$\lambda_{ij}^{P}=\frac{1}{2}(\eta^S_{ij}-(\eta^S)_{ij}^*)$ and,
finally, we take $\lambda_{ij}^{S}=\lambda_{ij}\,cos\,\theta$ and
$\lambda_{ij}^{P}=i \lambda_{ij}\,sin\,\theta$. For the CP
parameter $\theta$ we take an intermediate value in our numerical
calculations.
\item We assume that the diagonal couplings $\lambda_{ii}$ are
aware of the lepton family and obey the hierarchy,
$\lambda_{\tau\tau}>\lambda_{\mu\mu}>\lambda_{ee}$, and the
off-diagonal couplings, $\lambda_{ij}, i\neq j$ are family blind
and universal, obeying the equality $\lambda_{ij}=\kappa
\lambda_{ee}$ where $\kappa < 1$. Here we choose $\kappa=0.5$ and
take the numerical values of diagonal couplings so that the
experimental current limits of charged lepton EDMs can be reached.
\end{itemize}
Now, we start to predict the magnitudes of the charged lepton EDMs
and to estimate the appropriate range for the numerical value of
the scaling dimension $d_u$ by using the current experimental
limits of the EDMs. Here, we consider that the CP violating phase
is carried by the unparticle-charged lepton couplings so that the
charged lepton EDM arises with the unparticle mediation.

\begin{figure}[htb]
\vskip -3.0truein \centering \epsfxsize=6.8in
\leavevmode\epsffile{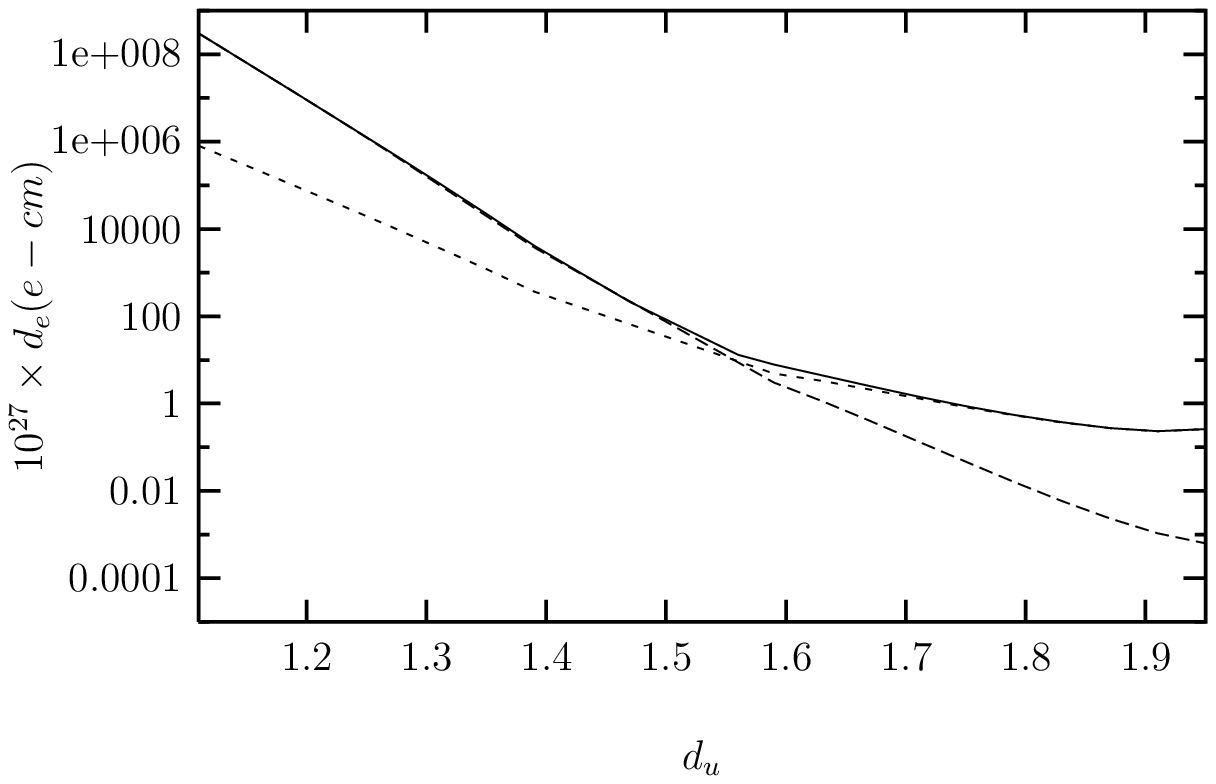} \vskip -3.0truein
\caption[]{$d_{e}$ with respect to the scale parameter $d_u$ for
the energy scale $\Lambda_u=10\, TeV$, the coupling
$\lambda_{ee}=0.01$ and the intermediate value of the CP parameter
$sin\theta=0.5$. Here the solid (dashed, small dashed) line
represents the EDM for total (diagonal, off-diagonal) part.}
\label{edmeScalardu}
\end{figure}

In  Fig.\ref{edmeScalardu}, we present the contribution of the
scalar unparticle to the EDM $d_{e}$ with respect to the scale
parameter $d_u$ for the energy scale $\Lambda_u=10\, TeV$, the
coupling $\lambda_{ee}=0.01$ and the intermediate value of the CP
parameter $sin\theta=0.5$. Here the solid (dashed, small dashed)
line represents the EDM for total (diagonal, off-diagonal) part.
It is observed that the there is a strong sensitivity to the scale
$d_u$ and the decrease of $d_u$ causes a sharp enhancement in the
EDM $d_e$. The diagonal (off-diagonal) part becomes strong for the
small (large) values of $d_u$ and the total contribution reaches
the experimental current value for the range of the scale $d_u$,
$1.6\leq d_u \leq 1.8$.

\begin{figure}[htb]
\vskip -3.0truein \centering \epsfxsize=6.8in
\leavevmode\epsffile{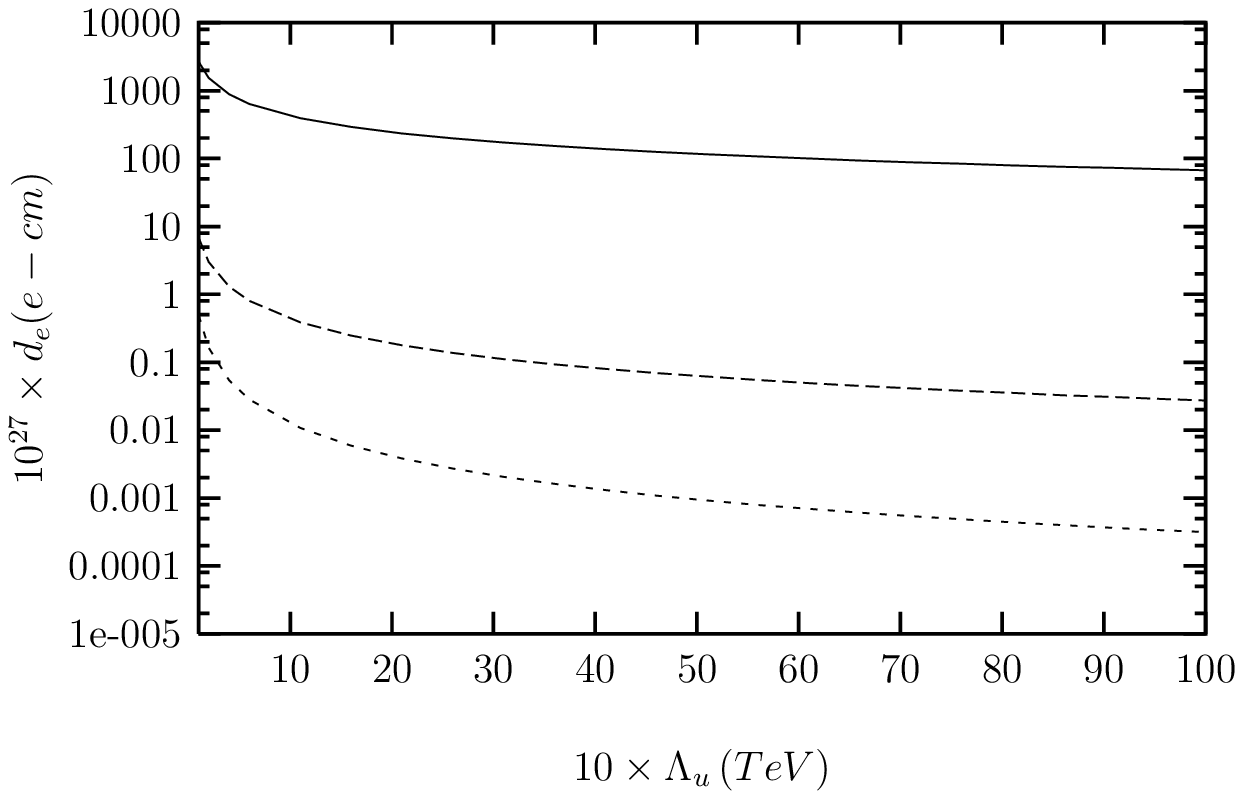} \vskip -3.0truein
\caption[]{$d_e$ with respect to the energy scale $\Lambda_u$, for
the coupling $\lambda_{ee}=0.01$ and the intermediate value of the
CP parameter $sin\theta=0.5$. Here the solid (dashed, small
dashed) line represents the EDM for
$d_u=1.4,\,d_u=1.6,\,d_u=1.8$.} \label{edmeScaEtau}
\end{figure}

Fig.\ref{edmeScaEtau} is devoted to the EDM $d_e$ with respect to
the energy scale $\Lambda_u$, for the coupling $\lambda_{ee}=0.01$
and the intermediate value of the CP parameter $sin\theta=0.5$.
Here the solid (dashed, small dashed) line represents the EDM for
$d_u=1.4,\,d_u=1.6,\,d_u=1.8$, including diagonal and off diagonal
terms. With the increasing value of the energy scale $\Lambda_u$
the EDM decreases and its sensitivity becomes weak, as expected.

\begin{figure}[htb]
\vskip -3.0truein \centering \epsfxsize=6.8in
\leavevmode\epsffile{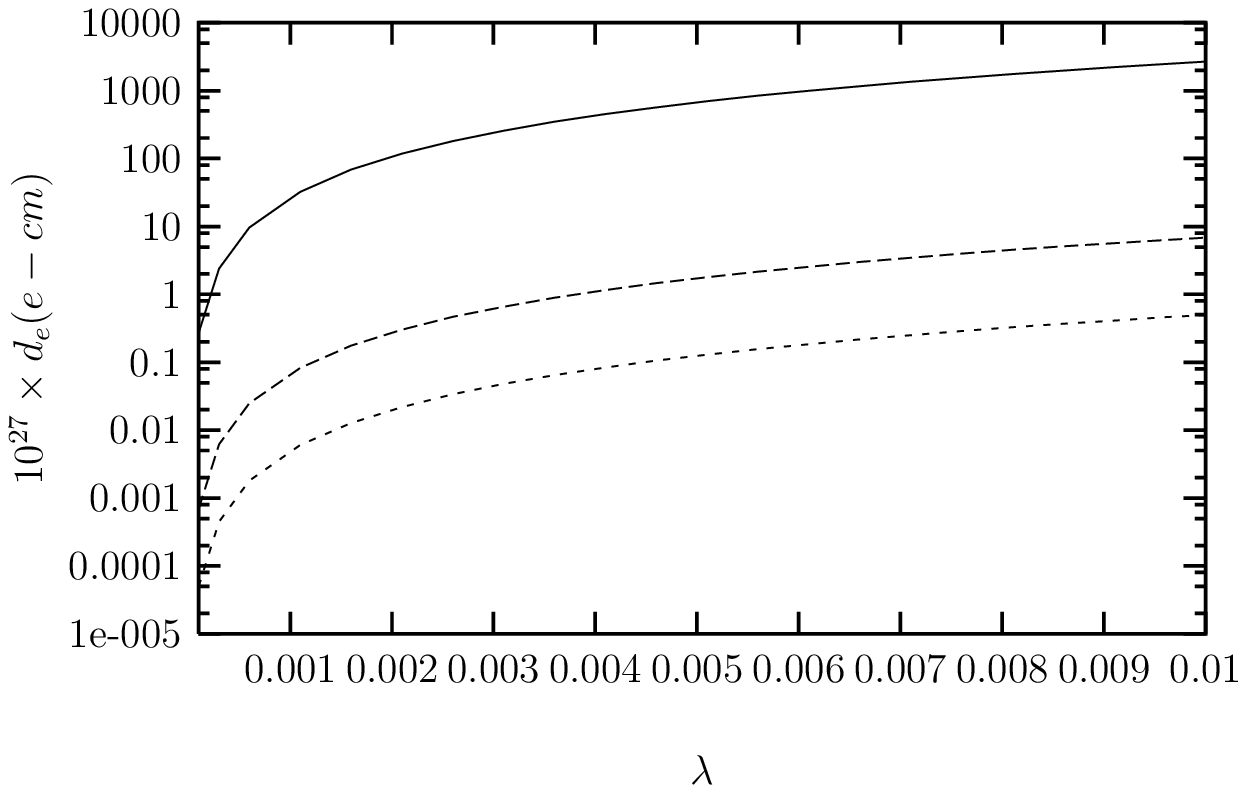} \vskip -3.0truein
\caption[]{$d_e$ with respect to the coupling $\lambda$ for the
energy scale $\Lambda_u=10\, TeV$ and the intermediate value of
the CP parameter $sin\theta=0.5$. Here the  solid (dashed, small
dashed) line represents the EDM for
$d_u=1.4,\,d_u=1.6,\,d_u=1.8$.} \label{edmeScaYukawa}
\end{figure}

Fig.\ref{edmeScaYukawa} represents the EDM $d_e$ with respect to
the coupling $\lambda=\lambda_{ee}$ for the energy scale
$\Lambda_u=10\, TeV$ and the intermediate value of the CP
parameter $sin\theta=0.5$. Here the solid (dashed, small dashed)
line represents the EDM for $d_u=1.4,\,d_u=1.6,\,d_u=1.8$,
including diagonal and off diagonal terms. This figure shows that
the EDM is strongly sensitive to the coupling $\lambda$ as
expected and, even for its small values of the order of $10^{-4}$,
it is possible to reach the experimental current limit, for the
scale dimension $d_u\sim 1.4$.

\begin{figure}[htb]
\vskip -3.0truein \centering \epsfxsize=6.8in
\leavevmode\epsffile{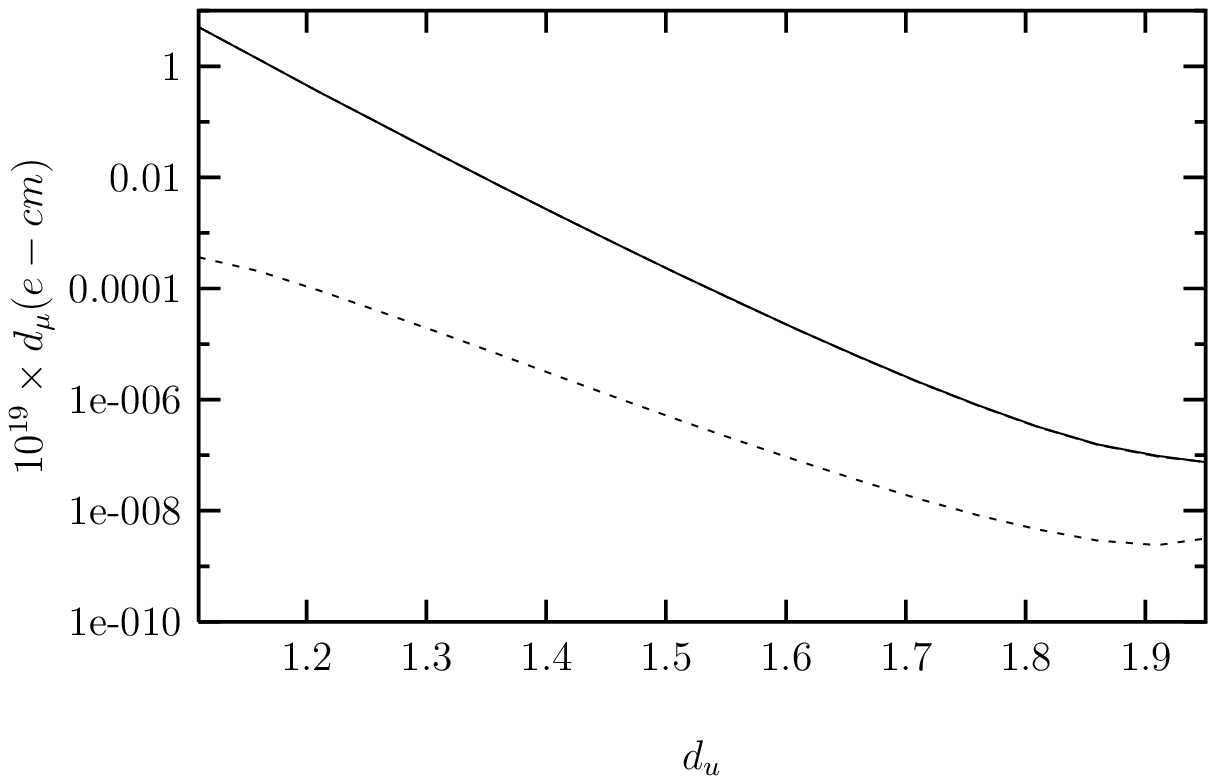} \vskip -3.0truein
\caption[]{The same as Fig. \ref{edmeScalardu} but for $d_{\mu}$
and the coupling $\lambda_{\mu\mu}=0.1$ } \label{edmmuScalardu}
\end{figure}

Fig.\ref{edmmuScalardu} is devoted to the contribution of the
scalar unparticle to the EDM $d_{\mu}$ with respect to the scale
parameter $d_u$ for the energy scale $\Lambda_u=10\, TeV$, the
coupling $\lambda_{\mu\mu}=0.1$ and the intermediate value of the
CP parameter $sin\theta=0.5$. Here the solid (dashed, small
dashed) line represents the EDM for total (diagonal, off-diagonal)
part. The EDM $d_\mu$ lies in the broad range $10^{-26}\, e-cm\leq
d_\mu \leq 10^{-19}\, e-cm$ for the interval of the scale
parameter $1.0< d_u < 2.0$ and the experimental current limit is
reached for the small values of the scale $d_u$, $d_u \leq 1.1$.

\begin{figure}[htb]
\vskip -3.0truein \centering \epsfxsize=6.8in
\leavevmode\epsffile{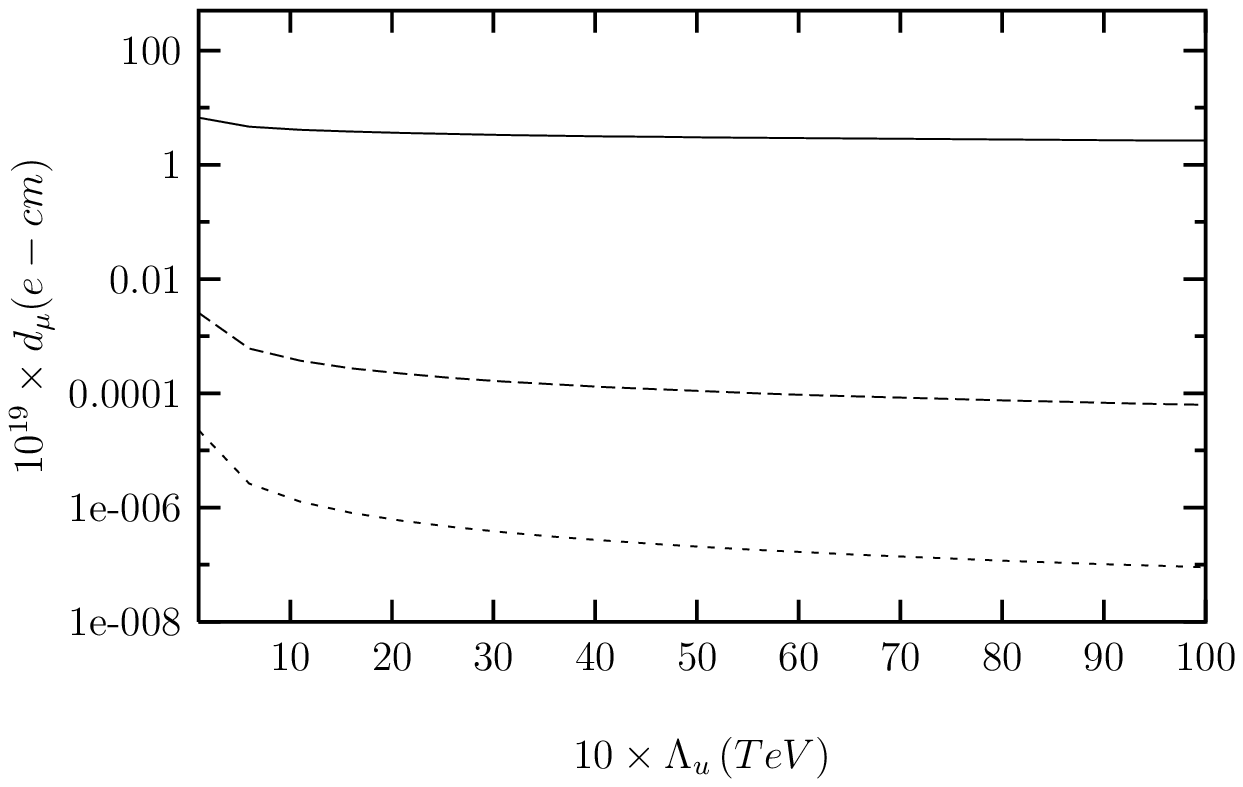} \vskip -3.0truein
\caption[]{$d_\mu$ with respect to the energy scale $\Lambda_u$,
for the coupling $\lambda_{\mu\mu}=0.1$ and the intermediate value
of the CP parameter $sin\theta=0.5$. Here the solid (dashed, small
dashed) line represents the EDM for
$d_u=1.1,\,d_u=1.4,\,d_u=1.6$.} \label{edmmuScaEtau}
\end{figure}

Fig. \ref{edmmuScaEtau} represents the EDM $d_\mu$ with respect to
the energy scale $\Lambda_u$, for the coupling
$\lambda_{\mu\mu}=0.1$ and the intermediate value of the CP
parameter $sin\theta=0.5$. Here the solid (dashed, small dashed)
line represents the EDM for $d_u=1.1,\,d_u=1.4,\,d_u=1.6$,
including diagonal and off diagonal terms. The increasing value of
the energy scale $\Lambda_u$ causes that the EDM decreases with
the suppressed sensitivity, similar to the electron EDM case.

\begin{figure}[htb]
\vskip -3.0truein \centering \epsfxsize=6.8in
\leavevmode\epsffile{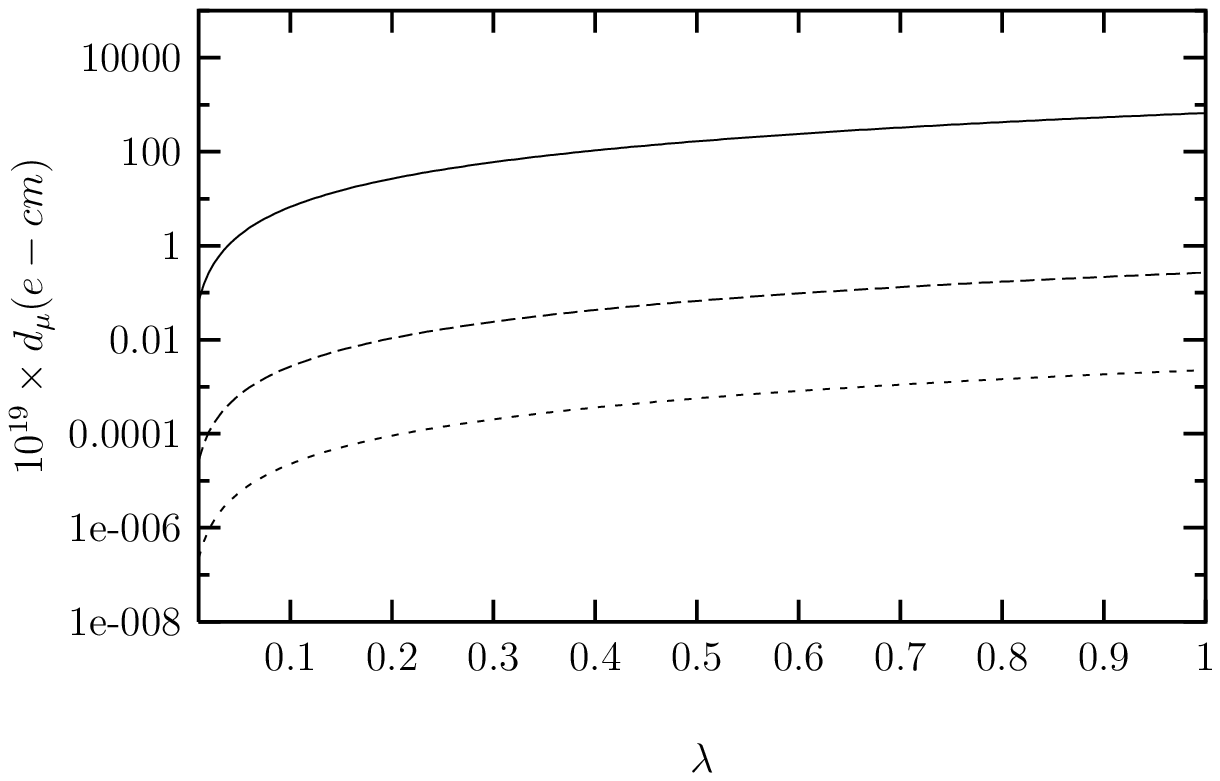} \vskip -3.0truein
\caption[]{$d_\mu$ with respect to the coupling $\lambda$ for the
energy scale $\Lambda_u=10\, TeV$ and the intermediate value of
the CP parameter $sin\theta=0.5$. Here the solid (dashed, small
dashed) line represents the EDM  for
$d_u=1.1,\,d_u=1.4,\,d_u=1.6$.} \label{edmmuScaYukawa}
\end{figure}

In Fig. \ref{edmmuScaYukawa} we show the EDM $d_\mu$ with respect
to the coupling $\lambda=\lambda_{\mu\mu}$ for the energy scale
$\Lambda_u=10\, TeV$ and the intermediate value of the CP
parameter $sin\theta=0.5$. Here the  solid (dashed, small dashed)
line represents the EDM  for $d_u=1.1,\,d_u=1.4,\,d_u=1.6$,
including diagonal and off diagonal terms. It is observed that the
experimental current limit can be reached if the Yukawa coupling
$\lambda$ is of the order of $0.1\, (1.0)$, for the scale
dimension $d_u < 1.1\,(d_u \sim 1.4)$.

\begin{figure}[htb]
\vskip -3.0truein \centering \epsfxsize=6.8in
\leavevmode\epsffile{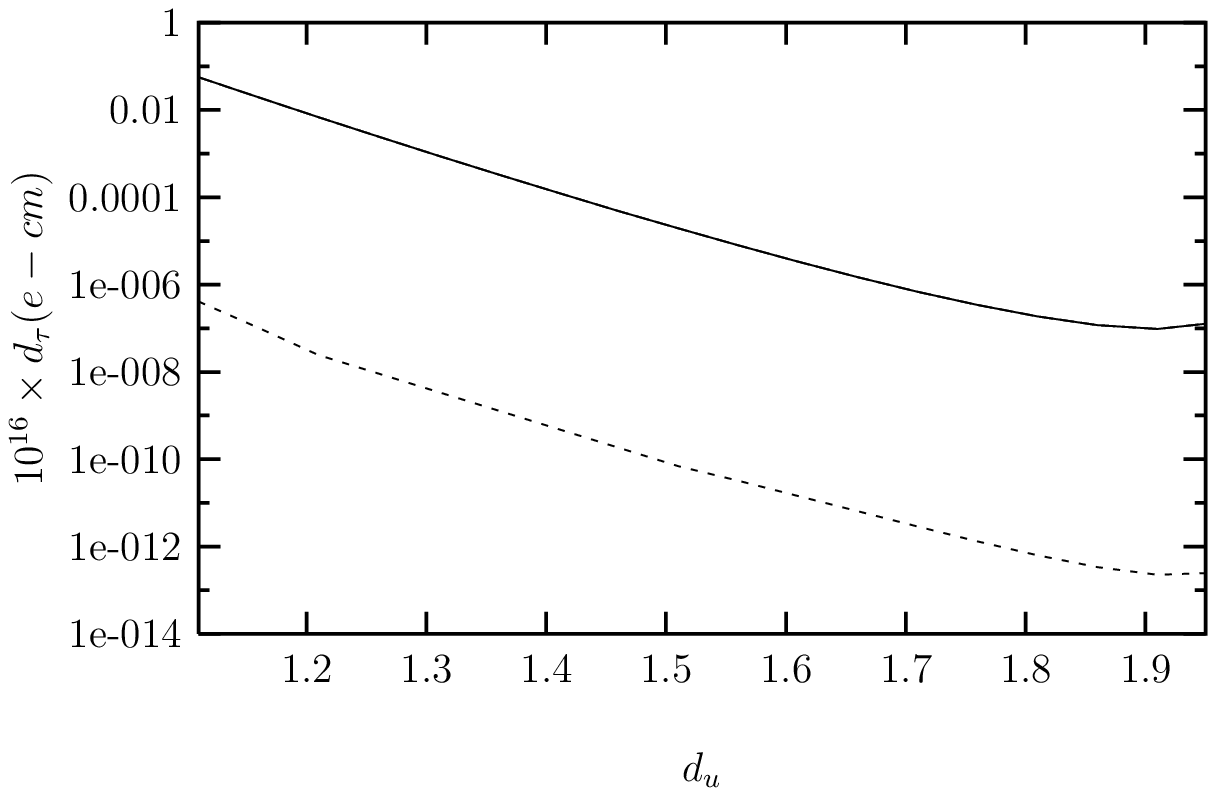} \vskip -3.0truein
\caption[]{The same as Fig. \ref{edmeScalardu} but for $d_{\tau}$
and the coupling $\lambda_{\tau\tau}=1.0$.} \label{edmtauScalardu}
\end{figure}

Figs. \ref{edmtauScalardu} is devoted to the contribution of the
scalar unparticle to the EDM $d_{\tau}$ with respect to the scale
parameter $d_u$ for the energy scale $\Lambda_u=10\, TeV$, the
coupling $\lambda_{\tau\tau}=1.0$ and the intermediate value of
the CP parameter $sin\theta=0.5$. Here the solid (dashed) line
represents the EDM for total (off-diagonal) contribution. For the
interval of the scale parameter $1.0< d_u < 2.0$ the EDM $d_\tau$
takes the values in the range $10^{-23}\, e-cm\leq d_\mu \leq
10^{-17}\, e-cm$. The off diagonal contribution  is almost five
order smaller than the diagonal contribution which coincides with
the total one. Here, the experimental current value can be reached
for the small values of the scale $d_u$, $d_u < 1.1$.

\begin{figure}[htb]
\vskip -3.0truein \centering \epsfxsize=6.8in
\leavevmode\epsffile{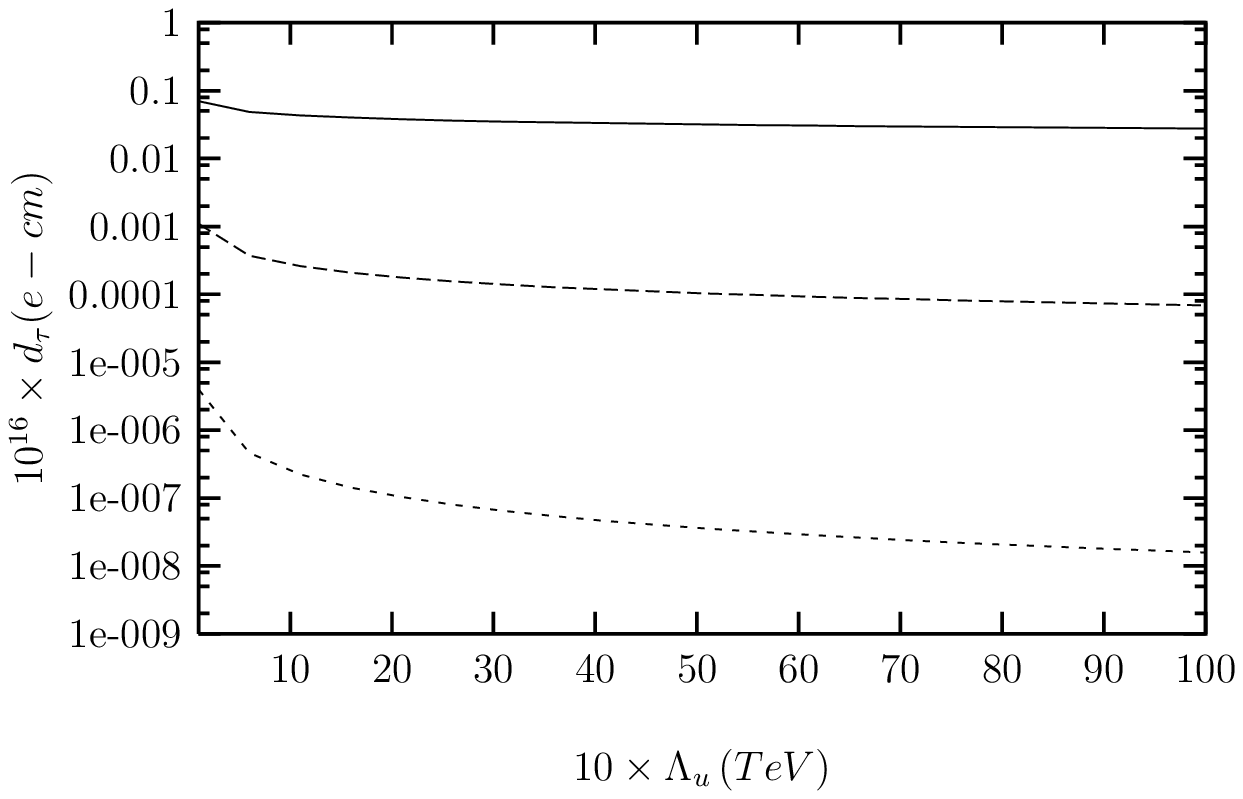} \vskip -3.0truein
\caption[]{$d_\tau$ with respect to the energy scale $\Lambda_u$
for the coupling $\lambda_{\tau\tau}=1.0$ and the intermediate
value of the CP parameter $sin\theta=0.5$. Here the solid (dashed,
small dashed) line represents the EDM  for
$d_u=1.1,\,d_u=1.4,\,d_u=1.6$.} \label{edmtauScaEtau}
\end{figure}

\begin{figure}[htb]
\vskip -3.0truein \centering \epsfxsize=6.8in
\leavevmode\epsffile{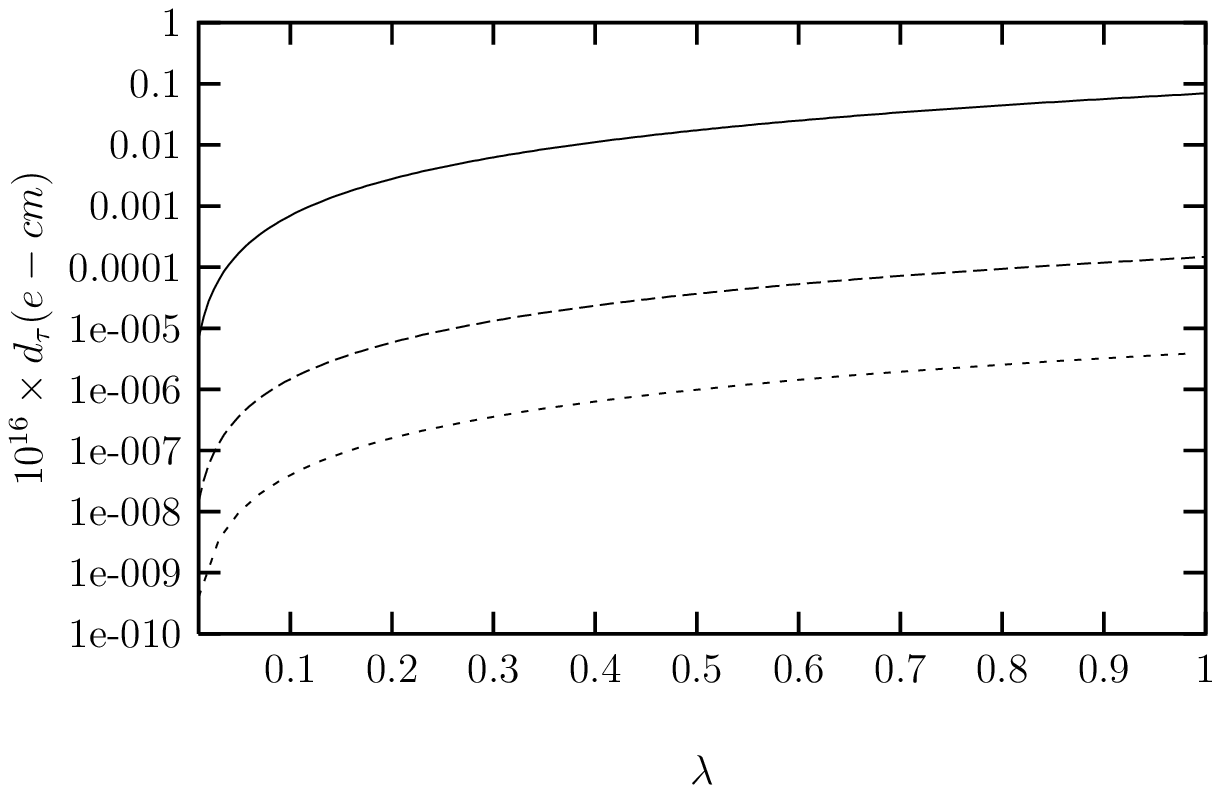} \vskip -3.0truein
\caption[]{$d_\tau$ with respect to the coupling $\lambda$ for the
energy scale $\Lambda_u=10\, TeV$ and the intermediate value of
the CP parameter $sin\theta=0.5$. Here the solid (dashed, small
dashed) line represents the EDM  for
$d_u=1.1,\,d_u=1.4,\,d_u=1.6$.} \label{edmtauScaYukawa}
\end{figure}

Fig. \ref{edmtauScaEtau}; \ref{edmtauScaYukawa} represents the EDM
$d_\tau$ with respect to the energy scale $\Lambda_u$; the
coupling $\lambda=\lambda_{\tau\tau}$, for the coupling
$\lambda_{\tau\tau}=1.0$; for the energy scale $\Lambda_u=10\,
TeV$ and the intermediate value of the CP parameter
$sin\theta=0.5$. Here the solid (dashed, small dashed) line
represents the EDM  for $d_u=1.1,\,d_u=1.4,\,d_u=1.6$, including
diagonal and off diagonal terms. Here it is observed that the
experimental value $\sim 10^{-16}\, GeV$ is reached for the energy
scale $\Lambda_u$, $\Lambda_u< 10\,TeV$, $d_u < 1.1$ and for the
coupling $\lambda$ at least of the order of $1.0$. The
contribution of the scalar unparticle to the EDM decreases with
the increasing values of the energy scale $\Lambda_u$.

\begin{figure}[htb]
\vskip -3.0truein \centering \epsfxsize=6.8in
\leavevmode\epsffile{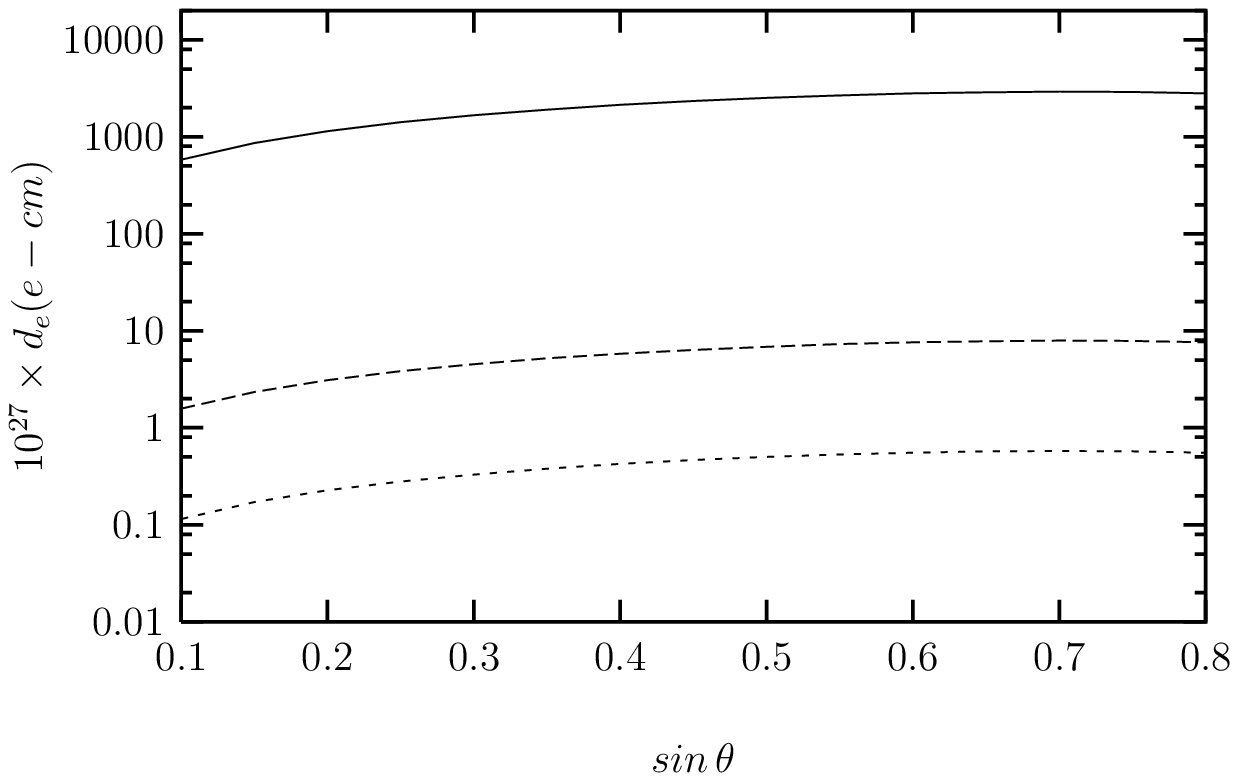} \vskip -3.0truein
\caption[]{$d_e$ with respect to the CP parameter $sin\theta$, for
the energy scale $\Lambda_u=10\, TeV$ and the coupling
$\lambda=0.01$. Here the solid (dashed, small dashed) line
represents the EDM  for $d_u=1.4,\,d_u=1.6,\,d_u=1.8$.}
\label{edmeScasintet}
\end{figure}

\begin{figure}[htb]
\vskip -3.0truein \centering \epsfxsize=6.8in
\leavevmode\epsffile{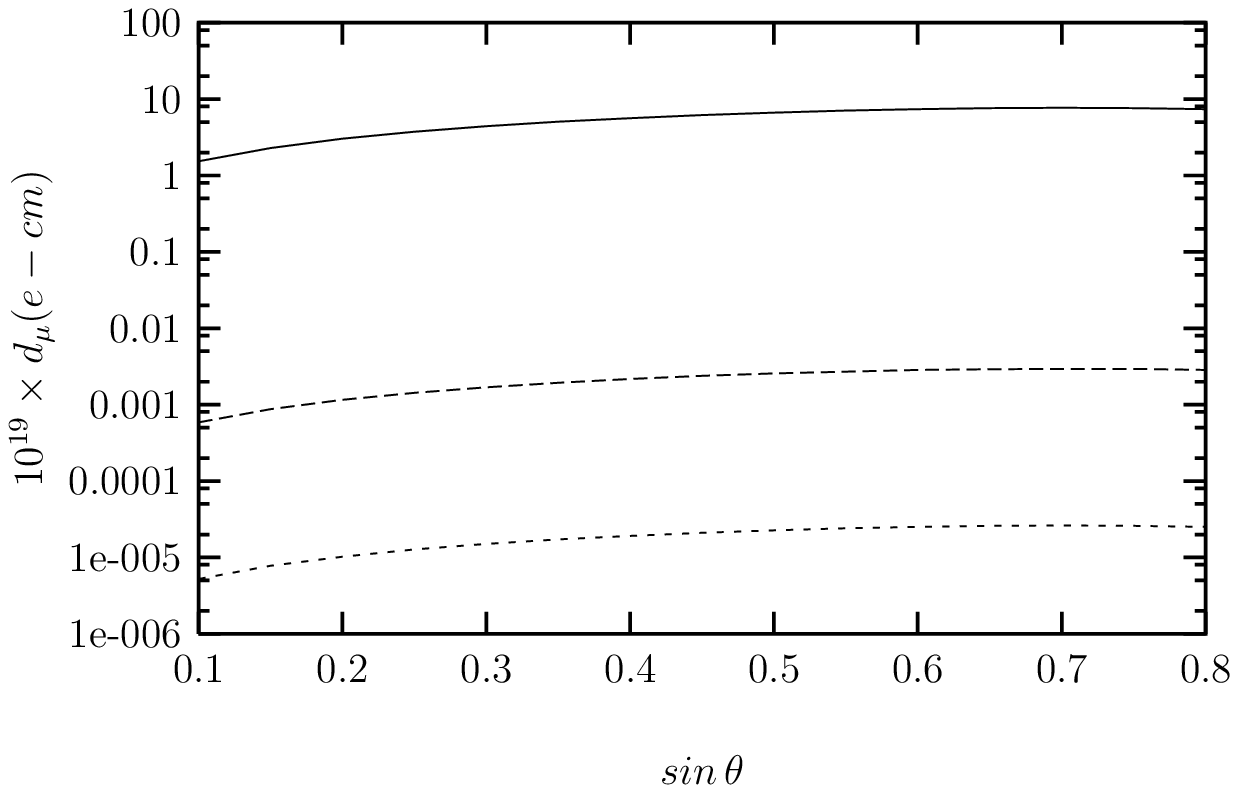} \vskip -3.0truein
\caption[]{$d_\mu$ with respect to the CP parameter $sin\theta$,
for the energy scale $\Lambda_u=10\, TeV$ and the coupling
$\lambda=0.1$. Here the solid (dashed, small dashed) line
represents the EDM  for $d_u=1.1,\,d_u=1.4,\,d_u=1.6$.}
\label{edmmuScasintet}
\end{figure}

\begin{figure}[htb]
\vskip -3.0truein \centering \epsfxsize=6.8in
\leavevmode\epsffile{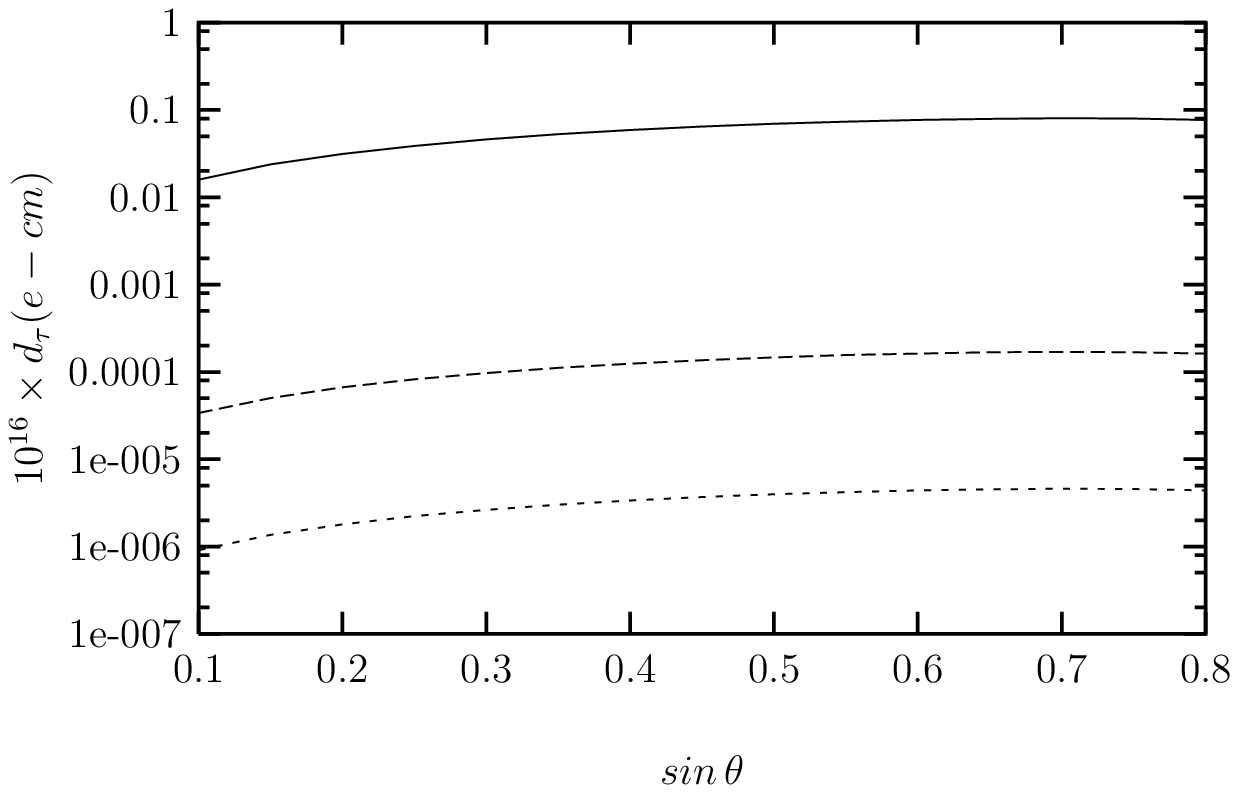} \vskip -3.0truein
\caption[]{The same as the Fig.\ref{edmmuScasintet} but for
$d_\tau$ and the coupling $\lambda=1$.} \label{edmtauScasintet}
\end{figure}

Finally, in Figs. \ref{edmeScasintet}; \ref{edmmuScasintet};
\ref{edmtauScasintet} we present the EDMs $d_e$; $d_\mu$; $d_\tau$
with respect to the CP parameter $sin\theta$, for the energy scale
$\Lambda_u=10\, TeV$ and the couplings $\lambda_{ee}=0.01$;
$\lambda_{\mu\mu}=0.1$; $\lambda_{\tau\tau}=1.0$, respectively.
Here the solid (dashed, small dashed) line represents the EDM  for
$d_u=1.4,\,d_u=1.6,\,d_u=1.8$; $d_u=1.1,\,d_u=1.4,\,d_u=1.6$;
$d_u=1.1,\,d_u=1.4,\,d_u=1.6$, including diagonal and off diagonal
terms. These figures show that the EDM changes almost one order in
the range $0.1<sin\theta <0.8$ and, even for the tiny CP
parameter, for the appropriate values of the scaling dimension, it
is possible to reach the current experimental limits.

Now we would like to summarize our results:

\begin{itemize}
\item The charged lepton EDMs are strongly sensitive to the
scaling dimension $d_u$. The experimental current limit of $d_e$
can be reached in the range $1.6\leq d_u \leq 1.8$. For $d_\mu$
and $d_\tau$ the current limits are reached for the small values
of the scale $d_u$, $d_u \leq 1.1$. The more reliable range for
the the scaling dimension can be estimated with more accurate
measurements of the charged lepton dipole moments.
\item  The contribution of the scalar unparticle to the EDM
decreases with the increasing values of the energy scale
$\Lambda_u$ as expected and the e sensitivity is large for the
numerical values $\Lambda_u<10\,TeV$.
\item The EDMs  are strongly sensitive to the coupling $\lambda$
and with the choice of the appropriate numerical value of the
scaling dimension, the experimental current limits of charged
lepton EDMs can be reached, even with tiny couplings.
\item The CP parameter $\theta$ is the source of the charged
lepton EDM and it is another free parameter that should be
restricted. The EDMs change almost one order in the range
$0.1<sin\theta <0.8$.
\end{itemize}

The forthcoming most accurate experimental measurement of charged
lepton EDMs would be illuminating to search the effects of
unparticles and to predict their scaling dimensions.
\newpage
\newpage

\begin{thebibliography}{1}
%
\bibitem{Georgi1}H. Georgi, {\it Phys. Rev.  Lett.}  {\bf 98}, 221601 (2007).
%
\bibitem{Georgi2} H. Georgi, {\it Phys. Lett.} {\bf B650}, 275 (2007).
%
\bibitem{Cheung1} K. Cheung, W. Y. Keung and T. C. Yuan, {\it Phys. Rev.  Lett.}
{\bf 99}, 051803 (2007).
%
\bibitem{Luo1}
M. X. Luo and G. H. Zhu, hep-ph/0704.3532.
%
\bibitem{Stephanov} M. A. Stephanov, Phys.\ Rev.\  D {\bf 76}, 035008
(2007).
%
\bibitem{Zwicky} R. Zwicky, hep-ph/0707.0677.
%
\bibitem{Cheung2} K. Cheung, W. Y. Keung and T. C. Yuan, hep-ph/0706.3155.

%
\bibitem{Luo2} M. X. Luo, W. Wu and G. H. Zhu, hep-ph/0708.0671.
%
\bibitem{Chen1} C. H. Chen and C. Q. Geng,  hep-ph/0705.0689.
%
\bibitem{Chen2}  C. H. Chen and C. Q. Geng, {\it Phys. Rev.} {\bf D76},
036007 (2007).
%
\bibitem{Chen3} C. H. Chen and C. Q. Geng, hep-ph/0709.0235.
%
\bibitem{SChen}  S. L. Chen and X. G. He, hep-ph/0705.3946.
%
\bibitem{Aliev1} T. M. Aliev, A. S. Cornell and N. Gaur,
hep-ph/0705.1326.
%
\bibitem{Aliev2} T. M. Aliev, A. S. Cornell and N. Gaur, {\it JHEP}
{\bf 07}, 072 (2007).
%
\bibitem{Aliev3} T.M. Aliev, M. Savci  hep-ph/0710.1505
%
\bibitem{Ding1} G. J. Ding and M. L. Yan, hep-ph/0705.0794.
%
\bibitem{Ding2} G. J. Ding and M. L. Yan, hep-ph/0706.0325.
%
\bibitem{Li1} X. Q. Li and Z. T. Wei, Phys. Lett. {\bf B651}, 380
(2007).
%
\bibitem{Li2} X. Q. Li and Z. T. Wei, hep-ph/0707.2285.
%
\bibitem{Liao1} Y. Liao, hep-ph/0705.0837.
%
\bibitem{Liao2} Y. Liao, hep-ph/0708.3327.
%
\bibitem{Liao3} Y. Liao and J. Y. Liu, hep-ph/0706.1284.
%
\bibitem{Fox} P. J. Fox, A. Rajaraman and Y. Shirman, hep-ph/0705.3092.
%
\bibitem{Catterall} S. Catterall and F. Sannino, {\it Phys. Rev.} {\bf D76},
034504 (2007).
%
\bibitem{Lu} C. D. Lu, W. Wang and Y. M. Wang, hep-ph/0705.2909.
%
\bibitem{Greiner} N. Greiner, hep-ph/0705.3518.
%
\bibitem{Choudhury} D. Choudhury, D. K. Ghosh and Mamta, hep-ph/0705.3637
%
\bibitem{Davou} H. Davoudiasl, hep-ph/0705.3636.
%
\bibitem{Chen4} S. L. Chen, X. G. He and H. C. Tsai, hep-ph/0707.0187.
%
\bibitem{Mathews} P. Mathews and V. Ravindran, hep-ph/0705.4599.
%
\bibitem{Zhou} S. Zhou, hep-ph/0706.0302.
%
\bibitem{Foadi} R. Foadi, M. T. Frandsen, T. A. Ryttov and F. Sannino,
hep-ph/0706.1696.
%
\bibitem{Bander} M. Bander, J. L. Feng, A. Rajaraman and Y. Shirman,
hep-ph/0706.2677.
%
\bibitem{Rizzo} T. G. Rizzo, hep-ph/0706.3025.
%
\bibitem{Goldberg} H. Goldberg and P. Nath, hep-ph/0706.3898.
%
\bibitem{Kikuchi} T. Kikuchi and N. Okada, hep-ph/0707.0893.
%
\bibitem{Giri1} R. Mohanta and A. K. Giri, hep-ph/0707.1234.
 %
\bibitem{Giri2} R. Mohanta and A. K. Giri, hep-ph/0707.3308.
%
\bibitem{Huang} C. S. Huang and X. H. Wu, hep-ph/0707.1268.
%
\bibitem{Krasnikov} N. V. Krasnikov, hep-ph/0707.1419.
%
\bibitem{Lenz} A. Lenz, hep-ph/0707.1535.
%
\bibitem{Ghosh} D. Choudhury and D. K. Ghosh, hep-ph/0707.2074.
%
\bibitem{Zhang} H. Zhang, C. S. Li and Z. Li, hep-ph/0707.2132.
%
\bibitem{Nakayama} Y. Nakayama,  hep-ph/0707.2451.
%
\bibitem{Deshpande1} N. G. Deshpande, X. G. He and J. Jiang,
hep-ph/0707.2959.
%
\bibitem{Deshpande2} N. G. Deshpande, S. D. H. Hsu and J. Jiang,
hep-ph/0708.2735.
%
\bibitem{Espinosa} A. Delgado, J. R. Espinosa and M. Quiros, hep-ph/0707.4309.
%
\bibitem{Neubert} M. Neubert, hep-ph/0708.0036.
%
\bibitem{Hannestad} S. Hannestad, G. Raffelt and Y. Y. Y. Wong,
hep-ph/0708.1404.
%
\bibitem{Das1} P. K. Das, hep-ph/0708.2812.
%
\bibitem{Das2} S. Das, S. Mohanty and K. Rao, hep-ph/0709.2583.
%
\bibitem{Bhattacharyya} G. Bhattacharyya, D. Choudhury and D. K. Ghosh,
hep-ph/0708.2835.
%
\bibitem{Majumdar} D. Majumdar, hep-ph/0708.3485.
%
\bibitem{AlanPak} A. T. Alan and N. K. Pak, hep-ph/0708.3802.
%
\bibitem{Freitas} A. Freitas and D. Wyler, hep-ph/0708.4339.
%
\bibitem{Gogoladze} I. Gogoladze, N. Okada and Q. Shafi, hep-ph/0708.4405.
%
\bibitem{Hur} T. i. Hur, P. Ko and X. H. Wu, hep-ph/0709.0629.
%
\bibitem{Anchordoqui} L. Anchordoqui and H. Goldberg, hep-ph/0709.0678.
%
\bibitem{Majhi} S. Majhi, hep-ph/0709.1960.
%
\bibitem{McDonald} J. McDonald, hep-ph/0709.2350.
%
\bibitem{Kumar} M. C. Kumar, P. Mathews, V. Ravindran and A. Tripathi,
hep-ph/0709.2478.
%
\bibitem{Cheung3} K. M. Cheung, W. Y. Keung, T. C. Yuan, hep-ph/0710.2230
%
\bibitem{Kobakhidze} A. Kobakhidze, hep-ph/0709.3782
%
\bibitem{Ding3} G. J. Ding and M. L. Yan, hep-ph/0709.3435.
%
\bibitem{Balantekin} A.B. Balantekin  , K.O. Ozansoy,
hep-ph/0710.0028.
%
\bibitem{Commins} E. D. Commins et.al. {\it Phys. Rev. A} {\bf
50}, 2960 (1994).
%
\bibitem{Bailey} J. Bailey et al, {\it Journ. Phys.} {\bf G4},
345 (1978);
%
\bibitem{Groom} Particle Data Group, D. E. Groom et.al. {\it European
Phys. Journ.} {\bf C15}, 1 (2000).
%
\bibitem{Khiplovich1} I. B. Khriplovich,  {\it Yad. Fiz.}
{\bf 44}, 1019 (1986) ; ({\it Sov. J. Nucl. Phys.} {\bf 44}, 659
(1986).
%
\bibitem{Khiplovich2} I. B. Khriplovich, {\it Phys. Lett.} {\bf
B173}, 193 (1986).
%
\bibitem{Czarnecki1} A. Czarnecki and B.
Krause, {\it Acta Phys. Polon.} {\bf B28}, 829 (1997).
%
\bibitem{Czarnecki2} A. Czarnecki and B.
Krause, {\it Phys. Rev. Lett} {\bf 78}, 4339 (1997).
%
\bibitem{Bhaskar} B. Dutta, R. N. Mohapatra, {\it Phys. Rev.}
{\bf D68}, 113008 (2003).
%
\bibitem{Iltmuegam} E. Iltan, {\it Phys. Rev.} {\bf D64}, 013013 (2001).
%
\bibitem{IltanNonCom} E. Iltan, {\it JHEP} {\bf 065}, 0305 (2003).
%
\bibitem{IltanExtrEDM} E. Iltan, {\it JHEP} {\bf 0404}, 018 (2004).
%
\bibitem{IltanSplitEDM} E. Iltan, {\it Eur. Phys. J.}
{\bf C44}, 411 (2005).
%
\bibitem{IltanSplitHiggsLocalEDM}  E. Iltan,  {\it Eur. Phys. J.}
{\bf C51}, 689 (2007).
%
\bibitem{IltanRSEDM} E. Iltan, hep-ph/0708.3765.
%
%
\end{thebibliography}
\end{document}